\documentclass[twocolumn,prl]{revtex4}
\usepackage{graphicx}
\newcommand{\be}{\begin{equation}}
\newcommand{\ee}{\end{equation}}
\newcommand{\bea}{\begin{eqnarray}}
\newcommand{\eea}{\end{eqnarray}}

\newcommand{\la}{\langle}
\newcommand{\ra}{\rangle}

\newcommand{\lp}{\left(}
\newcommand{\rp}{\right)}

\renewcommand{\phi}{\varphi}
\renewcommand{\epsilon}{\varepsilon}
\renewcommand{\vec}[1]{{\bf #1}}

\begin{document}

\title{Spin Filtered Edge States and Quantum Hall Effect in Graphene}
\author{Dmitry A. Abanin, Patrick A. Lee, Leonid S. Levitov}
\affiliation{
 Department of Physics,
 Massachusetts Institute of Technology, 77 Massachusetts Ave,
 Cambridge, MA 02139}

\begin{abstract}
Electron edge states 
in graphene in the Quantum Hall effect regime can carry both charge 
and spin. We show that spin splitting of the zeroth Landau level
gives rise to
counterpropagating modes
with opposite spin polarization.
These chiral spin modes
lead to a rich variety of spin current states, depending
on the spin flip rate. A method to control the latter locally is proposed.
We estimate Zeeman spin splitting enhanced by 
exchange, and obtain a spin gap of a few hundred Kelvin.
\end{abstract}

\maketitle

A new electron system with low carrier density and high mobility
was recently realized in two-dimensional graphene~\cite{Novoselov04}.
By varying the carrier density with a gate one can explore a range of 
interesting states, 
in particular the anomalous quantum Hall effect~\cite{Novoselov05,Kim} (QHE).
In contrast to the well-known integer QHE in silicon MOSFETs~\cite{Klitzing}
the QHE in graphene occurs at half-integer multiples of $4$, the degeneracy due to spin and orbit. This has been called the half-integer QHE.
The unusually large Landau level spacing makes QHE
in graphene observable at temperatures of $100\,{\rm K}$ and higher. 

Here we explore the spin effects in graphene QHE. 
In the presence of Zeeman splitting transport in graphene is described by an unusual set of edge states which we shall call chiral spin edge states. 
These states are reminiscent of the ordinary QHE edge states~\cite{Halperin}, 
but can propagate in opposite directions for opposite spin polarizations.
(As shown in~\cite{Kane}, similar states can arise due to
spin-orbit coupling
in the absence of magnetic field. However,
the weakness of spin-orbital effects makes the corresponding spin gap
quite small.)
The chiral spin edge modes can be used
to realize an interesting spin transport regime, in which 
spin and charge currents can be controlled independently. 
Observation of these phenomena is facilitated by fairly large magnitude of 
the spin gap. 
The gap is enhanced 
due to electron correlation and exchange, and 
can reach a few hundred
Kelvin for realistic magnetic field. 

The half-integer QHE in graphene was interpreted in terms
of a quantum anomaly of the zeroth Landau level~\cite{Gusynin}. 
Alternatively, these properties are easily understood
from the edge states viewpoint, similar to the usual QHE. This was done is Ref.~\cite{Peres} 
using numerical treatment of the zigzag edge. Here we present a continuum description of the edge states, using the massless Dirac model~\cite{diVincenzo}
which provides a good approximation for 
Carbon $\pi$-electron band near its center. 
We reduce the problem to a one-dimensional Schr\"odinger equation 
with a potential which depends on the boundary type. By comparing the behavior for armchair and zigzag boundary,
we show that the energy spectrum properties near the edge are universal 
and imply the half-integer QHE.

\begin{figure}
\includegraphics[width=1.64in]{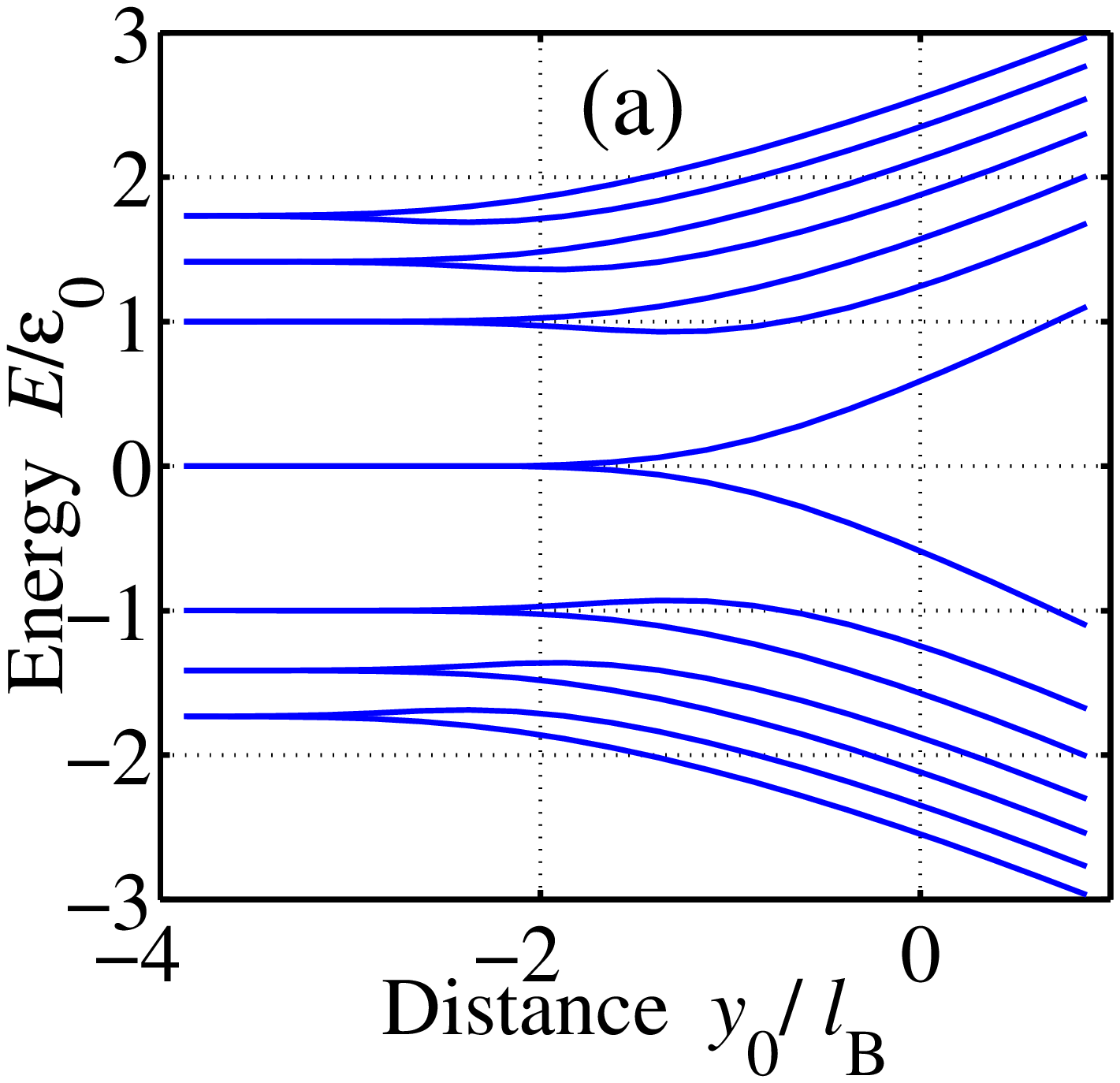}
\includegraphics[width=1.64in]{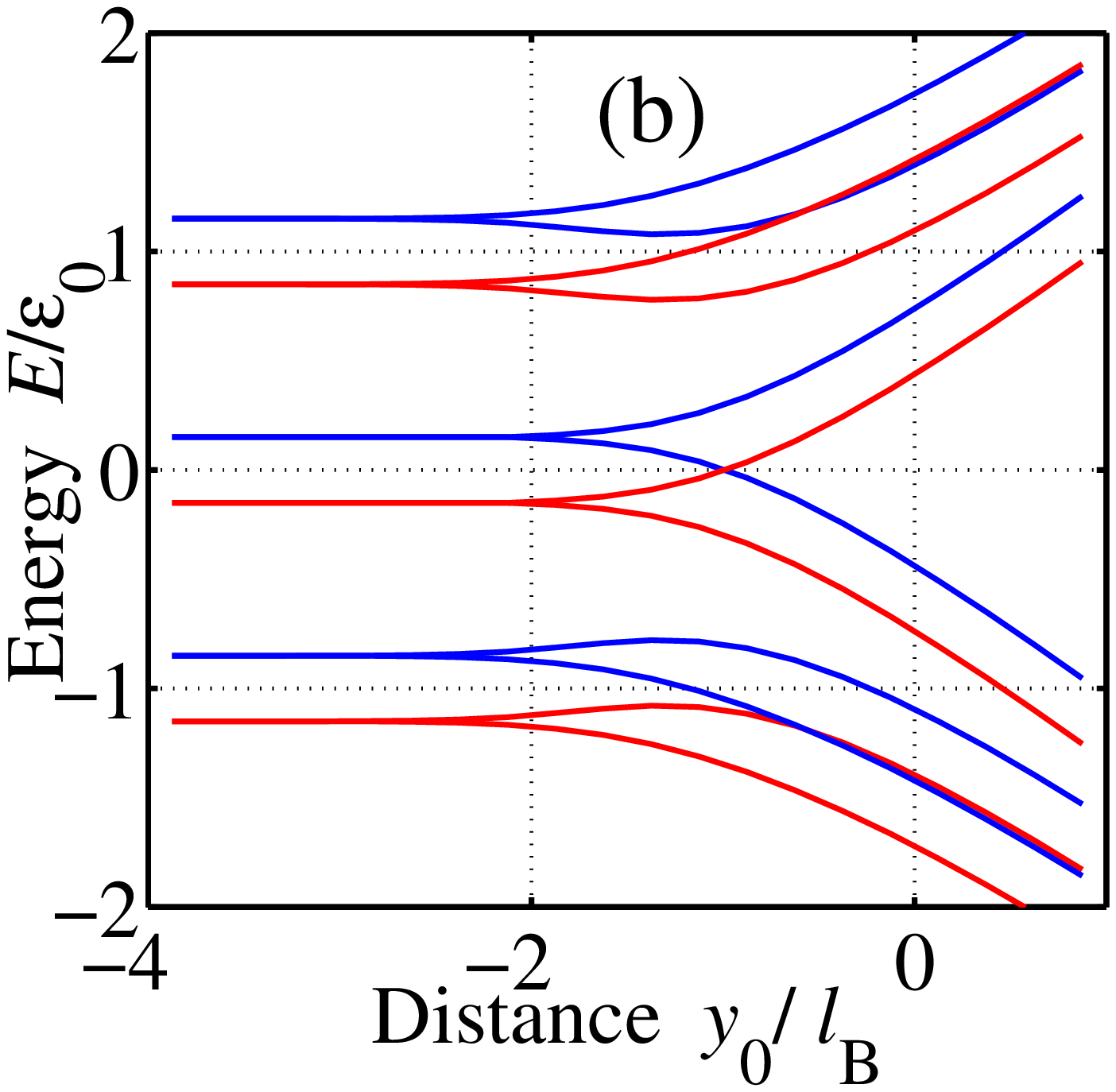}
\vspace{-3mm}
\caption[]{(a) Graphene energy spectrum near the armchair boundary
obtained from Dirac model, Eq.(\ref{eq:hamiltonian}). 
The boundary condition, Eq.(\ref{armchair_boundary}),
lifts the $K$, $K'$ degeneracy.
The odd integer numbers of edge modes 
lead to the half-integer QHE. 
(b) Spin-split graphene edge states: 
the blue (red) curves represent the spin up (spin down) states. 
These states propagate in opposite directions at zero energy.}
\label{fig1}
\vspace{-5mm}
\end{figure}

To interpret the half-integer QHE, let us inspect 
the energies of the first few Landau levels (LL) 
obtained for an armchair boundary (Fig.~\ref{fig1}(a)).
First we ignore electron spin. In the bulk the LL's are doubly degenerate, due to two species of Dirac particles located near $K$ and $K'$, the inequivalent 
corners of the first Brillouin zone. 
We note that the zeroth LL  
splits into two levels with positive and negative energies.
In contrast, the behavior of the edge states associated with
higher LL's is more conventional~\cite{Halperin}: the energies of positive (negative) LL's increase (decrease) 
as one approaches the boundary.
Hence in the spinless case the number of edge states
can take only odd integer values and
the Hall conductivity is odd integer in units of $e^2/h$. For example, when the chemical potential 
is between the $n=-1$ and $n=-2$ LL's, 
there are three branches of active edge states: 
two of them 
derived from the LL with $n=-1$ and one from the LL with $n=0$. 
As a result, although each Landau level filling factor is an integer,
the conductance at QHE plateaus is half-integer in units of $4e^2/h$
which accounts for both the $K$, $K'$ and spin degeneracy.

This behavior is modified in an interesting way by the spin splitting 
of LL's (Fig.~\ref{fig1}(b)).
When the chemical potential $\mu$ lies in the interval 
$-\frac12\Delta_s<\mu<\frac12\Delta_s$, the zeroth LL
is spin polarized, with only spin down states being filled.
However, there exists a branch of up-spin edge states going 
to negative energies. The states of this branch 
with $\epsilon<\mu$ will be filled and will contribute to transport
on equal footing with the down-spin states.
Notably, the up-spin and down-spin states have opposite
chiralities, i.e. they propagate in opposite directions. 
These states carry opposite charge currents but equal spin currents.
As a result,
the edge transport in the spin gap, $-\frac12\Delta_s<\mu<\frac12\Delta_s$, can be 
spin filtered.

The chiral spin edge states found here are similar 
to those predicted by Kane and Mele~\cite{Kane}, who considered
spin-orbit coupling in graphene
in the absence of Zeeman field. 
They obtain a spin gap which is
solely due to the spin-orbit coupling and 
is estimated as $\Delta _{SO}\sim 1\,{\rm K}$~\cite{Kane}. 
The small gap poses strong constraints on the amount of disorder
and on temperature, 
hindering experimental tests of the interesting predictions 
of Ref.~\cite{Kane}. 
In contrast, the magnetic field-induced gap is quite large:
the Zeeman energy of $\Delta _{s}=g\mu _B B\sim 15\, {\rm K}$ at $B\sim 10\, {\rm T}$ is further enhanced by exchange interaction,
putting it in an experimentally convenient range of hundreds of Kelvin.
Furthermore, 
one can alter the character of 
spin flip scattering at the edge to induce
backscattering among the spin polarized edge modes,
enabling 
experimental observation of a number of novel transport effects
(see below). 

To analyze the edge states, we
employ the Dirac model~\cite{diVincenzo} 
which describes the low-lying states as linear combinations
of four zero-energy Bloch functions with slow varying envelope functions 
$u_K$, $v_K$, $u_{K'}$, $v_{K'}$.
Here $u$ and $v$ correspond to the wave function components 
on two inequivalent atomic sites $A$, $B$, and the subscripts $K$, $K'$ 
denote the inequivalent Dirac points. 
The effective low-energy Hamiltonian, written near each of the
points $K$, $K'$ in terms of $u_{K,K'}$, $v_{K,K'}$, is of the form
\be\label{eq:hamiltonian}
H_K=v_0 
 \left[\begin{array}{cc}
         0 & \tilde{p}_+\\
         \tilde{p}_-& 0
      \end{array}
 \right], \,\,\,\,
H_{K'}=v_0
  \left[\begin{array}{cc}
         0 & \tilde{p}_- \\
         \tilde{p}_+ & 0 
 \end{array}\right], 
\ee
where $\tilde{p}_{\pm}=\tilde{p}_x\pm i\tilde{p}_y$,  
$\tilde{p}_{\mu}=p_{\mu}-\frac{e}{c}A_{\mu}$ and $v_0\approx 8\times 10^7\,{\rm cm/s}$ 
is Fermi velocity. 
The energy spectrum in the bulk is  
\be\label{eq:bulkLL}
E_n={\rm sgn }(n)|n|^{1/2}\epsilon_0
,\quad \epsilon_0=\hbar v_0\lp 2eB/\hbar c\rp^{1/2}
\ee
with integer $n$. For typical magnetic field  of $10\,{\rm T}$, 
the lowest LL separation
is estimated to be quite large, $\epsilon_0=E_1-E_0\approx 1000 \, {\rm K}$.
We now consider a graphene sample with an armchair edge parallel 
to the $x$ axis, using Landau gauge 
$A_x=-By$, $A_y=0$.
We eliminate the $v$ components 
of the wave function 
and consider the eigenvalue equations for the $u$ components: 
\bea\label{equations_uK_comp}
&& \frac12\left(p_y^2+(y-y_0)^2+1  \right)u_K=(E/\epsilon_0)^2 u_K, 
\\
\label{equations_uK'_comp}
&& \frac12\left(p_y^2+(y-y_0)^2-1\right)u_{K'}=(E/\epsilon_0)^2 u_{K'}, 
\eea
where $y_0=-p_x$. Here $x$, $y$ and $p_x$, $p_y$ are measured in the units
of $\ell_B=\left(\hbar c/eB\right)^{1/2}$ and $\hbar/\ell_B$.

To obtain the energy spectrum near the edge, 
one needs to supplement 
Eqs.(\ref{equations_uK_comp}),(\ref{equations_uK'_comp}) 
with suitable boundary conditions.
We obtain the latter from the tight-binding model
assuming that it is valid up to the very last row near the boundary. 
In the armchair case the boundary condition 
requires the wave function to vanish at both $A$ and $B$ lattice sites 
along the line $y=0$. For the envelope functions $u, v$,
taken at $y=0$, this means 
\be\label{armchair_boundary}
u_K=u_{K'},\quad v_K=v_{K'}
\ee
(see Ref.~\cite{McCann} for a more general analysis).

Instead of the problem in a halfplane 
with boundary conditions (\ref{armchair_boundary}),
it is more convenient to consider Eq.(\ref{equations_uK_comp}) 
on the negative half-axis $y<0$, and Eq.(\ref{equations_uK'_comp}) 
on the positive half-axis $y>0$. 
Then the first boundary condition in Eq.(\ref{armchair_boundary}) 
implies continuity of the wavefunction at $y=0$,
while the second condition 
implies  continuity of the derivative $\partial u/\partial y$.
Therefore the problem reduces to calculating energy levels in the potential
\be\label{eq:armchair_potential}
V(y)=\frac12(|y|-y_0)^2-\frac12{\rm sgn} (y),
\ee
defined on the entire line. 
Numerical solution of this problem
gives the energy dispersion 
shown in Fig.~\ref{fig1}.

We now briefly discuss the 
zigzag edge, 
another very common graphene edge type. 
In the absence of magnetic field 
this boundary supports a band of low-energy surface states~\cite{Fujita,Hatsugai}. 
Wavefunctions of these states decay away
from the boundary on a length scale of the order of lattice spacing, 
which is typically much shorter than the magnetic length $\ell_B$. 
Therefore, 
the surface states should not be sensitive 
to the magnetic field. This conclusion is supported 
by the numerical study~\cite{Peres}. 

In contrast to the surface states, special for zigzag edge, 
the LL behavior near the edge is universal and depends on the boundary type
only weakly. 
In the Dirac equation framework, 
the boundary condition for a zigzag edge is 
that either $A$ or $B$ components 
of the envelope function vanish at the boundary, 
depending on the orientation of the edge.
This yields two possible boundary conditions:
either $u_K=u_{K'}=0$ or $v_K=v_{K'}=0$ at the edge. 
Finding energy spectrum in this case 
requires solving Landau level problem with a hard-wall boundary condition, 
familiar from the usual QHE~\cite{Halperin}. 
The energy spectrum for the zigzag edge, 
given by the square root of the dispersion curves found in
Ref.~\cite{Halperin}, is qualitatively 
similar to the one in Fig.~\ref{fig1}. 
These results are consistent with the numerical analysis of
Ref.~\cite{Peres}.

Turning to the analysis of the spin gap,
let us consider the exchange energy for half-filled spin-degenerate
Landau level.
In the Hartree-Fock approximation, 
it is given by spin-dependent correlation
energy: 
\be
E=\sum_{i,j=+,-}
\frac12\int\int V(r-r')g_{ij}(r-r') d^2r\,d^2r'
\ee
with $g_{ij}(r-r')=\la n_i(r)n_j(r')\ra$ the pair correlation functions.
We consider Coulomb interaction of the form $V(\vec r)=e^2/\kappa r$
with the dielectric constant~\cite{Gonzalez}
describing screening due to the filled states,
$\epsilon<0$.

To describe correlations, 
we use the result for fully filled Landau level~\cite{Jancovici},
$g_0(r)=1-e^{- r^2/2\ell_B^2}$,
setting~\cite{Laughlin} 
\[
g_{++}(r)=n_+^2n_0^2g_0(r)
,\quad
g_{--}(r)=n_-^2n_0^2g_0(r)
,\quad
n_0=\frac{eB}{hc},
\]
where $n_{\pm}$ are the occupation fractions
of the up and down spin states, 
$n_++n_-=1$.
We model
the Coulomb interaction-induced correlations of particles with opposite spin, 
which are absent
for noninteracting electrons, by
$g_{+-}=n_+n_-n_0^2(1-\alpha e^{- r^2/2\ell_B^2})$, where $\alpha<1$
describes relative strength of Coulomb and exchange correlations.
This approach yields
\be
E_{\rm exchange}=-A n_+ n_-
,\quad
A=\lp\frac{\pi}{2}\rp^{1/2}\frac{e^2}{\kappa\ell_B}(1-\alpha).
\ee
To estimate spin polarization, we consider
Gibbs free energy per particle for the half-filled spin degenerate
Landau level, $G/N=E-TS$, where
\[
S=-T\sum_{i=\pm} n_i\ln n_i
,\quad
E=\frac12 E_Z(n_+ - n_-) -A n_+ n_-,
\]
with $E_Z$ the Zeeman energy.
The ground state, determined by $\delta G=0$ 
with $n_++n_-=1$,
satisfies
\be
T\ln(n_+/n_-)-E_Z+A(n_--n_+)=0.
\ee
In the absence of exchange interaction, $A=0$, this would give 
$n_-/n_+=e^{-E_Z/T}$. The more realistic case of exchange $A$ large 
compared to $E_Z$ can be analyzed by setting $E_Z=0$.
In this case we have a phase transition at $T_c=\frac12 A$.
At low temperature, $T\ll T_c$, the concentrations $n_\pm$ satisfy
$n_-/n_+=e^{-A/T}$, i.e. the spin gap at $T\ll T_c$ can be estimated
as $\Delta = A$. 

To obtain numerical value of $\Delta$, we use the RPA estimate
of the screening function~\cite{Gonzalez}, with $e^2/\hbar v\approx 2.7$:
\be
\kappa = 1+ 2\pi e^2\cdot\frac1{4\hbar v} \approx 5.24.
\ee
%
Comparing to the LL separation 
$\epsilon_0$, Eq.(\ref{eq:bulkLL}),
%
\be
\Delta =\frac{\pi^{1/2}e^2}{2\kappa\hbar v } (1-\alpha)\epsilon_0
\approx 0.456\cdot (1-\alpha)  \epsilon_0.
\ee
Note that $\Delta$ is proportional to the Dirac LL energy $\epsilon_0$ and scales as $B^{1/2}$, in contrast to the Zeeman energy. If we 
 use $\alpha=0$, i.e. ignore correlations 
of electrons with opposite spins, we obtain $\Delta\simeq 450\, {\rm K}$ for $B\simeq 10\,{\rm T}$. (This approximation, while
giving correct order of magnitude, 
may somewhat overestimate the exchange contribution.)

\begin{figure}
\includegraphics[width=2.6in]{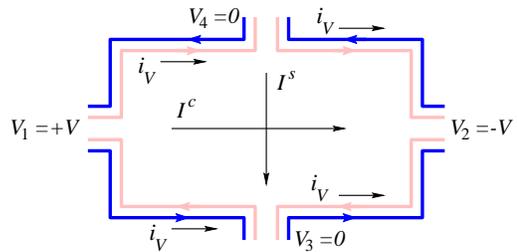}
\vspace{-3mm}
\caption[]{Four-terminal device 
with chiral spin edge states (no backscattering). 
In response to charge current 
of $I^c=2i_V=4e^2V/h$  between reservoirs 1 and 2, 
pure spin current of $I^s=2i_V$ flows
between 4 and 3, while Hall voltage is zero.}
\label{fig3}
\vspace{-1mm}
\end{figure}
\begin{figure}
\includegraphics[width=3.2in]{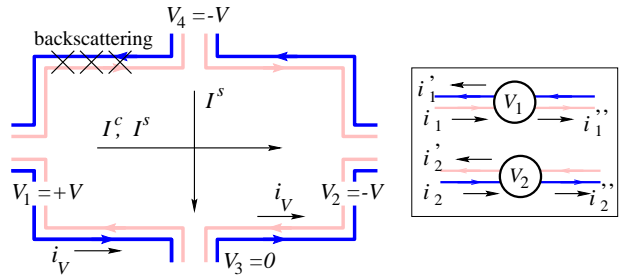}
\vspace{-3mm}
\caption[]{
Same as in Fig.~\ref{fig3} with backscattering between edge states
induced by altering local spin-flip rate. Strong backscattering 
gives rise to asymmetric current flow with the current between 1 and 2
fully spin polarized. Inset shows how Hall voltage probe 
can be used to detect spin current.
}
\label{fig4}
\vspace{-3mm}
\end{figure}

We now discuss possible experimental tests
of the chiral spin edge states, using the four-terminal device 
shown in Figs.~\ref{fig3},\ref{fig4}. We assume that each 
contact injects 
both spin polarizations
with the same voltage $V_k$, i.e. full spin mixing takes place in the leads. 
The {\it charge} current flowing out of the $k$-th contact is given by 
%
\be\label{eq:charge_current}
I_k^c=\sum _{k'} g_{kk'}\left(V_k-V_{k'} \right),         
\ee
where $g_{kk'}$ is the Landauer conductance of the edge channel connecting 
contacts $k$ and $k'$, and $V_k$ are voltages on the contacts. 
In the presence of backscattering among the edge states,
the conductance $g_{kk'}$ can be expressed in terms of 
transmission matrix $T_{kk'}$ of the edge channels~\cite{Buttiker},
with $g_{kk'}=T_{kk'}e^2/h$. 
The voltage probes are defined by the condition $I_k^c=0$ and Eq.(\ref{eq:charge_current}) is solved for the current and voltage at each contact~\cite{Buttiker}. 

The {\it spin current} is then determined as follows.
The spin current flowing from the reservoir $k$ to reservoir $k'$ 
is given by the sum of 
outgoing current of $e^2V_k/h$, 
reflected current $(1-T_{kk'})e^2V_k/h$ and the current 
from $k'$-th reservoir, $T_{kk'}e^2V_{k'}/h$. We obtain
$I_{kk'}^s=\pm\left[(2-T_{kk'})V_k+T_{kk'}V_{k'}\right]e^2/h$
(the three contributions to the spin current are of the same sign,
plus if the channel $k\to k'$ is spin up and minus if it is spin down).
Thus the total spin current flowing out of the $k$-th contact is
\be\label{eq:spin_current}
I_k^s=\sum_{k'}  
I_{kk'}^s =\sum_{k'} \epsilon_{kk'} g_{kk'} (V_k-V_{k'}), 
\ee
where $\epsilon_{kk'}=-\epsilon_{k'k}$ equals $+1$ ($-1$) 
when the current from $k$ to $k'$ is carried by spin up (spin down) electrons. 


The general relations (\ref{eq:charge_current}), (\ref{eq:spin_current}) 
become more transparent in two simple limiting cases, illustrated 
in Figs.~\ref{fig3},~\ref{fig4}. 
First we inspect the clean limit 
with no backscattering at the edge (Fig.~\ref{fig3}).
Applying voltage of $2V$ between contacts 1 and 2 we obtain pure 
{\it spin current} across the sample equal to $2i_V$, where $i_V=e^2V/h$. 
There is also charge current of $2i_V$ between 1 and 2 and no Hall voltage, 
yielding $\rho _{xx}=h/2e^2$.

Alternatively, let us introduce 
strong backscattering in the shoulder 
connecting 1 and 4, 
such that the resistance between 1 and 4 is infinite, 
while all the other shoulders are clean (Fig.~\ref{fig4}). 
Such a system acts as a ``spin filter''. Indeed, the current flowing from 1 to 2 is {\it spin polarized}. 
The Hall voltage in this case is nonzero,
equal to the half of the voltage between 1 and 2. 

Interestingly, the Hall voltage measures spin current rather 
than charge current. This is illustrated by Fig.~\ref{fig4} inset
which depicts currents $i_1,\, i_2$ that flow into the voltage probes, 
equilibrate and flow out
($i_{1,2}'=\frac12 i_{1,2}$). The Hall voltage is proportional to the incoming {\it spin} 
current:
\be
\label{eq:hallvoltage}
V_1-V_2=\frac12(i_1-i_2)h/e^2=-I_s\cdot h/2e^2.
\ee
Such voltage probes can be attached 
to the top or bottom leads in Fig.{~\ref{fig3},~\ref{fig4}} to verify the presence of spin currents. (We caution that currents $i_1,\, i_2$ are 
affected by the probes and should be 
calculated self-consistently using (\ref{eq:charge_current}).) 

We now discuss spin-flip scattering mechanisms. 
One can make qualitative observations which 
(a) show that spin-flip scattering is strongly suppressed and 
(b) suggest a way to manipulate it.  
We assume that the main source of the spin-flip scattering 
is spin-orbit (SO) interaction, which includes the intrinsic term proportional to $\sigma _{z}$ and the Rashba term~\cite{Kane}. 
The former interaction is ineffective 
when electron spins are polarized along the $z$ axis, 
amounting merely to an energy shift 
of the order of 
$\Delta_{SO} \sim 1\, {\rm K}$, 
without the up-down spin mixing.
At the same time, the Rashba term turns out to be extremely small, 
$\lambda_{R}\sim 0.5\, {\rm mK}$~\cite{Kane}, making spin-flips negligible. 

The situation changes,
however, when an {\it in-plane magnetic field} is applied.
Now the 
intrinsic SO interaction 
admixes the spin projections along the tilted field, leading to 
a small avoided crossing of the chiral spin edge states
of magnitude $\Delta _{SO}$. 
Using a local gate one can detune from 
this crossing 
by $\Delta E\gg \Delta_{SO}$. 
In this case,
the spin-flip scattering probability 
is estimated 
as $\Delta_{SO}^2/\Delta E^2$, which gives a factor of $10^{-2}$ for $\Delta E \sim 10\, {\rm K}$. 
The spin-flip scattering rate will be further reduced if the disorder 
at the edge has a short length scale compared to $\ell_B$
(recent STM experiments~\cite{Kobayashi,Niimi} report $a\sim 1\,{\rm nm}$). 
For typical magnetic length, $\ell_B\sim 10\,{\rm nm}$ at $B\sim 10\, {\rm T}$,  this gives spin-flip scattering suppressed by a factor of $a^2/l_B^2\sim 10^{-2}$. 
These observations show that backscattering at the edge should be 
strongly suppressed and suggest that by gating the QHE edge
in the presence of in-plane magnetic field
one can control 
local spin-flip rates independently. In particular, by tuning $\Delta E$ to zero  in one shoulder of the four terminal device, 
one could realize the situation depicted in Fig.~\ref{fig4}. 


Up to now we have focused on the LL spin splitting and ignored the orbital $K,K'$ degeneracy. 
However, the exchange effects must lift the $K,K'$ degeneracy as well. 
Moreover,  the $K,K'$ exchange gap should be of the same size as the spin gap $\Delta$
since electron interaction has the same strength for different spin and valley species
($SU(4)$ symmetry).
The spontaneously appearing order parameter will have orientation fixed by
weak nonsymmetric interactions, the difference between the intra and inter-sublattice coupling for $K,K'$, and the Zeeman energy for spin splitting.
The spin filter effects discussed above assume that spin splitting occurs before orbital splitting.        

We conclude by addressing the experimental situation. 
Our theory predicts a plateau of $R_{xy}=0$ between $eV=\pm \frac12\Delta$ 
which was observed in a recent experiment~\cite{Zhang06}. 
This paper also reports full lifting of the four-fould degeneracy of the zeroth LL at high magnetic fields $B>10\,{\rm T}$ which is consistent with the 
predicted strength of $K,K'$ exchange. We attribute the suppression of the zeroth LL splitting at lower magnetic fields to the disorder present 
in the currently available samples~\cite{MacDonald}. 
The observed value of $R_{xx}$ at $B=25\,{\rm T}$ 
is about $40\,{\rm k}\Omega$. 
According to our picture, this means that the chiral edge states are not localized and transmission 
coefficients of the edge channels are about $1/3$. Thus a spin Hall current may already be present in this case. 

After the completion of this work, we became aware of Ref.~\cite{Fertig} 
where a continuum description of the edge states given 
in the first part of this paper is also reported. 

We acknowledge support by NSF grant number DMR-0517222 (PAL) and NSF-NIRT DMR-0304019 (LL).


\end{document}